# Comparative study of Discrete Wavelet Transforms and Wavelet Tensor Train decomposition to feature extraction of FTIR data of medicinal plants

Pavel Kharyuk[a,b], Dmitry Nazarenko[c,*], Ivan Oseledets[a,d]

[a]*Skolkovo Institute of Science and Technology, Skolkovo Innovation Center, Building 3, Moscow, Russia, 143026*
[b]*Faculty of Computational Mathematics and Cybernetics, Lomonosov Moscow State University, Leninskiye Gory 1-52, GSP-1, Moscow, Russia, 119991*
[c]*Faculty of Chemistry, Lomonosov Moscow State University, Leninskiye Gory 1-3, GSP-1, Moscow, Russia, 119991*
[d]*Institute of Numerical Mathematics of the Russian Academy of Sciences, Gubkina st. 8, Moscow, Russia, 119991*

## Abstract

Fourier-transform infra-red (FTIR) spectra of samples from 7 plant species were used to explore the influence of preprocessing and feature extraction on efficiency of machine learning algorithms. Wavelet Tensor Train (WTT) and Discrete Wavelet Transforms (DWT) were compared as feature extraction techniques for FTIR data of medicinal plants. Various combinations of signal processing steps showed different behavior when applied to classification and clustering tasks. Best results for WTT and DWT found through grid search were similar, significantly improving quality of clustering as well as classification accuracy for tuned logistic regression in comparison to original spectra. Unlike DWT, WTT has only one parameter to be tuned (rank), making it a more versatile and easier to use as a data processing tool in various signal processing applications.

## Highlights

- Wavelet Tensor Train (WTT) was tested as feature extraction technique for classification and clustering tasks

- Discrete Wavelet Transform (DWT) and WTT with various preprocessing techniques were compared in application to feature extraction from FTIR spectra

- "Contrasting" approach based on soft thresholding of decomposition coefficients before subtraction was proposed as a way to significantly improve quality of clustering

## 1. Introduction

Medicinal plants continue to remain important source for new drugs development in the field of pharmacology and Western medicine [1–3]. Complimentary and alternative medicine (CAM), on the other hand, seeks to provide relief through use of medicinal plants and medicinal plants based preparations [4]. Recent decades mark rapid expansion of herbal drug production [5]. Increasing attention to Traditional Chinese Medicine (TCM), where about 80% of formulae include plant material, also contributed to rising popularity of herbal medicines [6]. Standardization and quality control of such products and raw plant materials pose a considerable challenge due to complexity of chemical composition [7]. Nonetheless, research and clinical studies continue to be conducted, albeit with limited efficiency [8; 9]. "Fingerprint" approach became one of the most popular tools to tackle this analytical problem [8]. This approach relies on combination of highly-informative analysis methods (mass-, NMR- and IR-spectroscopy, chromatographic methods, etc.) and machine learning techniques [10]. Although chromatography [11; 12] and NMR-spectroscopy [13] also gained significant attention in fingerprinting approaches, vibrational spectroscopy is still predominant in the field [14; 15]. IR-spectroscopy is a fast and relatively low cost method, making it an attractive choice as instrumentation for both research and routine analysis of raw plant material and finished goods. In recent decades, IR data was extensively applied to geographical origin and species differentiation [14], quality control [16], plant age confirmation (important for ginseng species, for example) [17], discrimination of adulterated/authentic samples [18], etc. Most of above mentioned studies share similar and well established experimental design: collection of several samples per group, IR data acquisition, data preprocessing and application of machine learning technique. With the first step depending on the topic of particular study and the second being rather straightforward, machine learning step usually gets the most attention. Nevertheless, as inconspicuous as it may seem, data preprocessing may be crucial to the whole study. IR-spectra of complex samples reflect sum of spectral responses generated by individual compounds. Chemical data in such form is very hard to interpret by humans. Therefore, applications of spectroscopy to industrial and scientific tasks are often assisted by a variety of mathematical techniques [19]. Reasoning behind mathematical processing is, among other things, to separate relevant chemical information from the rest of the data and help answer analytical question at hand. Many vibrational spectroscopy variants (Near-infrared spectroscopy, NIR; mid-infrared spectroscopy, MIR; Raman spectroscopy, etc.) commonly applied to medicinal plant analysis share a similar batch of accepted preprocessing techniques and approaches. Smoothed spectra, first and second derivatives of spectra as well as truncated regions of original spectra are commonly used in plant analysis as input data for machine learning algorithms [20]. Alternatively, application of wavelet transform can be a useful way to preprocess IR-data. Overall similarity of major bands in spectra implies, that relevant chemical information is hidden in minor (high frequency) bands. Wavelet approaches are exactly tailored for searching such bands in data [21] and are widely used for signal [22; 23] and image processing [24].

---

*Corresponding author

In case of discrete wavelet decompositions, filters are predetermined and do not change according to the specifications of given data. Recently, adaptive wavelet-like filtering based on tensor decomposition which can be a better alternative was proposed [25] and later applied for data compression [26; 27]. In this study, discrete wavelet transforms and wavelet tensor train decomposition were applied to mid-IR spectral data of medicinal plant extracts to explore their merits as feature extraction techniques in machine learning pipelines for classification and clustering.

## 2. Materials and methods

### 2.1. Chemicals and plant material

Ethanol was purchased from Merck (Germany). Deonized water was purified with Milli-Q water system (Millipore, Milford, MA, USA). Plant material was partly obtained from botanists and partly from commercial suppliers. Plant species and their quantities are listed in Table 1.

### 2.2. Sample extraction and IR experiment

1 g of each plant was powdered in agate mortar and sonicated for 1 hour in 10 mL of 70% EtOH. 2 mL of crude extract were centrifuged for 10 min (10000 g); supernatant was dried in vacuum centrifuge. Dry extract was mixed with KBr in mortar and formed into tablets. Spectra were recorded in the mid-IR range between 4000 and 400 cm$^{-1}$. Parts from $2000 - 400$ cm$^{-1}$ range were used in all computational experiments. Resulting spectra from our dataset are displayed on Figure 1.

### 2.3. Classification and clustering

Species identification task of unknown samples as well as exploratory analysis for dataset of analyzed samples may be formulated in terms of machine learning problems. Classification techniques are applied for identification: being an example of supervised learning where parameters of algorithm are tuned in order to minimize specific loss dependent on labels of training set, it generalizes inner properties of data in an effort to be able to separate input samples from different classes. On the other hand, in clustering task separation and grouping of samples are carried out without any prior information and are entirely based on inner structure of input data.

Crucial step in machine learning assisted data analysis is feature extraction task. Performance of algorithms is strongly dependent on the representation of data, and an appropriate feature space may drastically increase it. One of the widely used tools for obtaining valuable representation of data is wavelet decompositions. However, such representation by itself may not be sufficient to meet the challenge. For instance, data may contain noise which could be reduced with additional processing step via thresholding of wavelet coefficients. In general, any linear transformation $W$ of input $x$ is often followed by mapping it to another feature space with specific non-linearity. In book [28] an optimal nonlinear transform of wavelet features is presented for searching images by handwritten pictures:

$$f(Wx) = \text{sign}\left[\theta_\tau^{\text{hard}}(Wx)\right],\tag{1}$$

where hard thresholding $\theta_\tau^{\text{hard}}(\cdot)$, $\tau > 0$, serves as feature selector across wavelet coefficients, and sign function $\text{sign}(\cdot)$ performs simple quantization of the result. In our experience such choice is also adequate to be used in a FTIR classification pipeline.

However, in clustering individual aspects of samples move to the forefront, and another strategy is preferred. The basic idea is to remove some kind of common trend from data. Similar approach was used in [29] to extract individual features from images via matrix factorization with designed properties. In this work such concept is referred to in a generalized sense as *data contrasting*. In clustering analysis it was explored that relatively good results are given by using data that is contrasted in the following way:

$$x_{\text{contrasted}} = x - W^{-1}\left[\theta_\tau^{\text{soft}}(Wx)\right],\tag{2}$$

where $\theta_\tau^{\text{soft}}(\cdot)$ is a soft thresholding.

For convenience of the reader, clustering and classification algorithms used in the work are briefly described in Appendix.

## 2.4. Discrete Wavelet Transform

Wavelet transforms are designed to analyze signals with long low-frequency trends and fast occurrences of high frequency events. Wavelet decomposition makes it possible to localize both frequency and time (spatial) changes (however, with restrictions on resolution). Discrete wavelet transform (DWT) appears as a result of discretization of continuous wavelet transform which is a convolution of a signal with a specifically defined function $\psi(t)$ called mother wavelet:

$$W(\tau, s) = \frac{1}{\sqrt{s}} \int\limits_{-\infty}^{+\infty} f(t)\psi\left(\frac{t-\tau}{s}\right)dt = (f, \psi_{\tau,s})_{L_2},\tag{3}$$

where $\psi_{\tau,s} = s^{-1/2}\psi(\frac{t-\tau}{s})$, parameters $s \in \mathbb{R}\backslash\{0\}$ and $\tau \in \mathbb{R}$ stand for scaling and translation respectively. It has been shown that a number of discrete wavelet transforms are related to filtering of input signal by specific low-pass (defined by scaling function $\varphi(t)$; captures approximation of a signal) and high-pass (defined by mother wavelet $\psi(t)$; captures details) filters followed by down-sampling of results [30]:

$$\begin{aligned}
\hat{y}^{\text{low}}[n] &= (x * \varphi)[n] = \sum_k x[n]\varphi[n-k], \quad y^{\text{low}} = \hat{y}^{\text{low}} \downarrow 2, \\
\hat{y}^{\text{high}}[n] &= (x * \psi)[n] = \sum_k x[n]\psi[n-k], \quad y^{\text{high}} = \hat{y}^{\text{high}} \downarrow 2.
\end{aligned}\tag{4}$$

Different mother wavelets assign specific wavelet transforms. In this study 5 families of discrete wavelet decompositions were used, namely: Daubechies wavelets and its modification (symlets), coiflets, biorthogonal and reverse biorthogonal wavelets [31].

In practice, all signals have finite size, and it is necessary to extend them near borders to ensure perfect reconstruction of input with inverse transform. Different extension modes bring about so-called edge effects like discontinuities of signal or its derivatives. The influence of 7 padding schemes was studied in this work: zero, constant, symmetric, reflect, periodic, smooth, and periodization extension modes (see [32]).

It is also reasonable to perform multilevel transformation of signal where DWT is applied recursively to low-passed version of signal. Such scheme is usually referred to as filter bank. From this point of view one can define wavelet-like transformation through specifying filter bank with similar properties (like orthogonality, pseudosparseness of the result, and others). In the next subsection one of the techniques to generate such wavelet-like mappings is briefly described.

## 2.5. Wavelet Tensor Train decomposition

Wavelet Tensor Train (WTT) was introduced in [25] as the algebraic tool for generating adaptive wavelet-like orthogonal transformations. In this decomposition, input signal $x \in \mathbb{R}^{n_1 \cdot \ldots \cdot n_d}$ is to be virtually tensorized to form multi-dimensional array (tensor) $X \in \mathbb{R}^{n_1 \times \ldots \times n_d}$. As a multilevel DWT, WTT may be defined recursively: at first, unfolding matrix $A_{(k)}$, for the first dimension $k = 1$ through decomposition with SVD, where left singular matrix $U_k$ subsequently forms the first filter; then the first $r_k$ rows of the rest part of decomposition are taken and reshaped into new tensor $X_k$, to which considered operations are to be applied recursively to obtain filter bank $\Phi_{\text{WTT}}$:

$$
X_k = \begin{cases} \underset{n_1 \times n_2 \times \ldots \times n_d}{\text{Tens}}(x), & \text{if } k = 1, \\ \underset{(r_{k-1}n_k) \times n_{k+1} \times \ldots \times n_d}{\text{Tens}} \left[ \text{Cut}_{r_{k-1}\uparrow} \left( U_{k-1}^T A_{k-1} \right) \right], & \text{if } k > 1 \end{cases},
$$

$$
A_k = \text{unfold}_1 \left( X_k \right) = U_k \Sigma_k V_k^T,
$$

$$
\Phi_{\text{WTT}} = \{ U_i \}_{i=1}^{d-1},
$$

where $\text{Cut}_{r_k\uparrow}[X]$ is an operator that cuts first $r_k$ rows of matrix $X$, $\underset{n_1 \times \ldots \times n_d}{\text{Tens}}(x)$ is a tensorization (reshaping) of $x \in \mathbb{R}^{n_1 \cdot \ldots \cdot n_d}$ to $n_1 \times \ldots \times n_d$ shape, $\text{unfold}_k(X)$ means taking the k-mode unfolding matrix. Hyperparameters $(r_1, \ldots, r_{d-1})$ are called ranks of WTT decomposition, they control the size of each filter. As for the tensorization process, mode sizes were selected as $n_i = 2$, and cubic interpolation of signals was used to ensure that their resulting length is a degree of 2. As in the DWT approach, one may use padding schemes (for example, zero padding) as an alternative to interpolation; however, in this case it would be difficult to extend already learned algorithm to data sampled in a different way, for example when using equipment with different resolution.

Computed filters may be applied to any other input of appropriate size. All parts of signal that were dropped at $k$-th step are to be concatenated and saved as output of transform. It is worth mentioning that origin of WTT filters is not necessarily the same signal as one to be decomposed. In [26; 27] the idea of joint (group) filters was elaborated for lossy compression task: given set of equally sized $d$-dimensional objects was considered as a $(d+1)$-dimensional tensor, and filter computation procedure was performed for such stacked data. Then all filters except one for group axis were used for data compression. In this work it was studied how such common WTT transform performs in feature extraction task in comparison with discrete wavelets.

## 2.6. Cross-validation and parameter selection

For computational convenience, two-stage cross validation (CV) scheme was used. At first, grid search was performed to select best parameters for all classification algorithms by maximizing accuracy values in 4-fold CV with specified random state. For final comparison of best performing algorithms, 25 times repeated 4-fold CV (100 runs in total) with different random state was used.

Selection of parameters for agglomerative clustering was performed in the same way: selection of both clustering parameters (affinity, linkage) and hyperparameters of decompositions was done on dataset randomly splitted into 4 parts with algorithm being run on respective triplets. Best parameters were used for clustering of the full dataset.

Behavior of processing approaches was examined using 3 basic feature spaces: original signal ($f$), its first derivative ($f'$) and its second derivative ($f''$). For further processing both standard scaled and non-treated version were used, giving 2 additional parameters for grid search: centering and normalization by standard deviation. Another parameter was either taking absolute values or keeping (scaled) signals as is.

If signal was processed with DWT (or WTT), wavelet and padding mode (or rank) were to be selected, followed by choosing the type of thresholding, either hard or soft, and thresholding constant, $\tau$.

For logistic regression, two more parameters participated in selection procedure: penalty ($l_1$ or $l_2$) and inverse of regularizing constant ($C = \lambda^{-1}$). In clustering, changes in scores related to using contrasting were additionally inspected.

To evaluate performance of classification algorithms learned on different feature spaces, standard quality metrics were used: accuracy (fraction of correctly classified samples) and $F_1$ score (harmonic mean of precision, which is a fraction of relevant samples among all that were classified as current class, and recall, which is a fraction of correctly classified as current class samples among all samples from this class). As for clustering, three scores were inspected: Rand score, which is similar to accuracy, mutual information score, which is an information-theoretical distance measure between joint and product of marginal distributions, and Fowlkes-Mallows score, which is the geometric mean between precision and recall. The first two scores were corrected for chance (referred to as adjusted).

### 2.7. Computational tools

All the experiments were implemented in Python programming language. Anaconda Python distribution [33] has been used as programming framework, it includes various pre-built packages for scientific computations. In this study the following packages were used: numpy [34], scipy [35], pandas [36], scikit-learn [37], pywavelets [32], matplotlib [38], seaborn [39], statmodels [40].

All scripts and data acquired in the study can be found at GitHub repository: `https://github.com/kharyuk/chemfin-ftir`. Computational experiments are structured as Jupyter Notebooks [41].

## 3. Results and discussion

In this study, 7 medicinal plant species were selected to generate FTIR dataset with most of them (except for *Inula Helenium*) covered by Russian Pharmacopoeia. Linear discriminant analysis, logistic regression classifier, and agglomerative clustering were used to evaluate effects of different preprocessing steps and feature extraction approaches based on discrete wavelet transforms and wavelet tensor train decomposition. General schematic description of approaches covered by this study is summarized at Figure 2.

It is worth noting that in our study the dimensionality of feature space is higher than the number of samples. Although in general large dimensionality of feature space may lead to overfitting, especially for low-sampled datasets, redundant feature space is not always a curse, if it shares several additional properties. For instance, in the situation of overcomplete representation, learning feature space is to be purposefully enlarged in order to increase robustness in presence of noise, to enforce sparseness and to facilitate matching of data structure [42]. As a special class of overcomplete representations, sparse representations are useful for classifiers due to the better linear separability of distinct data in higher-dimensional spaces [43; 44]. Similar reasonings may be applied to large but sparse feature spaces, which can be achieved using wavelet decompositions with so-called vanishing moments followed by thresholding that filters out all small values - such processing makes the result sparse. As an additional measure to prevent severe overfitting, inspection of gap between quality values at training and test sets was monitored.

$F_1$ scores for classification were close to accuracy values (Tables 2, 3), which is to be expected from nearly balanced dataset. Thus, all results were presented and analyzed on the basis of accuracy analysis. Clustering results were observed in terms of three scores: adjusted Rand index, adjusted mutual information and Fowlkes-Mallows score. Due to high degrees of correlation between these scores (see Table 4), in our study the most pessimistic adjusted Rand score was selected as reference point.

### 3.1. Initial preprocessing

Dataset was utilized in three forms: original spectra, 1st and 2nd derivative spectra. Centering, scaling and taking absolute value were tested as initial preprocessing steps (Figure 3, 4). Taking absolute values concurrently with centering lead to decreases in classification accuracy in most cases, and standard scaling often increased accuracy (sometimes only scaling or centering works better). Same observations were valid for clustering results.

### 3.2. Linear Discriminant Analysis

In case of LDA, hard thresholding worked better than soft for DWT/WTT based approaches (Figure 3, (a)–(b)). If sign function was applied after it (Figure 3, (d)–(e)), it made algorithms less sensitive to type of thresholding and also reduced the gap between accuracy on train and test parts (i.e., helped to fight overfitting).

From Table 2 and Figure 3, (a)–(e) one can see that accuracy on test part is sometimes higher than on train - this is a problem of relatively small dataset which must get eliminated with increasing number of available samples. An additional point is that such behavior was observed in case of using sign quantization.

Even though sets of parameters selected with short CV demonstrated very promising results, large CV showed steep decrease in accuracy, especially for the original spectra. Such outcome was caused by averaging results from various random partitions of dataset. DWT had the best overall accuracy, followed by WTT with thresholding (without using sign function), and further below were WTT with sign function and original feature space. As for WTT, rank 1 showed the best accuracy on original and first derivative spectra, while rank 6 was the best for second derivative spectra. Nevertheless, low absolute accuracy values showed that even with tuned parameters WTT decomposition evidently was not compatible with LDA classification of second derivative spectra. Generally, the higher the rank, the easier it is to get overfitting.

For some cases LDA performs better on thresholded features without additional quantization (for instance, WTT on original signal and second derivative). However, there is no any significant gain in performance when WTT used with LDA classifier. Alongside with that, features extracted by WTT are valid for logistic regression: there are sets of parameters which maximize the accuracy for both WTT and DWT to the same extent.

Summarizing, feature extraction based on WTT/DWT decompositions coupled with non-linearity slightly elevated LDA efficiency. Nevertheless, LDA showed significant overfitting tendencies for original signals and their thresholded decompositions, relaxed by using quantization. Overall, tuned logistic regression significantly outperforms LDA if appropriate features were used.

### 3.3. Logistic Regression

In comparison to LDA, logistic regression algorithm requires selection of two more parameters related to regularization which was used as the counter-measure for overfitting. With that, grid search over various combinations of preprocessing steps and hyperparameters of decompositions for LR was performed. Original and derivative spectra without additional processing resulted in very poor accuracy measured in short CV (Figure 3, (f)) - 0.34 and 0.17 respectively.

For DWT, padding did not significantly influence final performance of LR (Figure 5, (a)-(e)), it is assumed that tuning of other parameters can compensate for it. Even there were no vivid regular patterns when selecting padding scheme, apparently, it may be reasonable to optimize it manually for each problem. Among other wavelet families, almost all coiflets worked better for derivative spectra than original spectra. In general, most wavelets worked better with first derivatve signals, with short gaps between train and test parts and consistently higher accuracy (Figure 3, (g)), while original and second derivative of signals took the 2nd and the 3rd places. At the same time, original spectra were significantly worse for coiflets and Daubechies wavelets of higher orders. Additionally, it was observed that tuning of parameters can make all results similar, except for some rare cases. The same states with thresholding, for both DWT and WTT, - it also has to be tuned manually. Regretfully, there is no way around it. On the other side, even search of threshold on a coarse grid can give good results.

As shown on Figure 5, (f), there is no clear dependency of performance on rank of decomposition for WTT, presumably because of the cumulative effect of other tuned parameters like regularization constant and threshold. According to Table 3 and Figure 3, (h), properly tuned WTT slightly outperformed DWT on original and first derivative spectra and lose on second derivative. One of the obvious arguments in favor of WTT is that sharing the similar accuracy to DWT, it has only one hyperparameter to tune, the rank. In case it is imperative to use second derivative spectra, coiflets could do worse as the initial choice. Accuracy on second derivative could also be elevated by using higher order approximations (in this study the 2nd order one was used).

### 3.4. Hierarchical Agglomerative Clustering

Apart from classification tasks, which are usually formulated when developing tools for automatization of routine analysis procedures, clustering is also frequently used, mainly in chemo-taxonomic studies of agricultural and medicinal plants. Therefore, it is also essential to examine influence of data transformations on clustering performance. Better clustering (in terms of adequate grouping of similar objects) can give more meaningful insights into scientific problems at hand.

From Figure 4, (a)–(c) it appears that clustering quality for signals without decomposition was substantially lower for most configurations, and relatively better performance was observed for 1st derivative of spectra with sample-wise scaling by standard deviation followed by taking absolute values of the result.

While best performance without decompositions was shown by first derivative spectra, for WTT/DWT processed signals best results were achieved on original spectra with contrasting and soft thresholding (Figure 4, (d)–(i); Figure 6). It is worth to note also that centering and scaling of feature space is preferred to sample-wise preprocessing in case of DWT/WTT decompositions. As seen from Figure 6, (f), WTT performance positively correlated with rank for all used feature spaces. Coiflets shared the same property as observed in classification with Logistic Regression: they work better with spectra derivatives (Figure 6, (b)). For DWT, there were no distinct dependencies between model parameters (wavelet family, padding scheme) and clustering quality.

Behavior of algorithms experienced minor changes when they were applied to full dataset (Table 4): DWT only slightly improved performance for derivative spectra and achieved better results for second derivative of spectra in comparison to original signals. WTT based approach, on the other hand, showed absolute best results in conditions of the experiment.

Hierarchical clustering with best performing configurations visualized in form of dendrograms plotted for original spectra ((a), (c), (e)) and their first derivative ((b), (d), (f)) is shown in Figure 7.

## 4. Conclusions

In this study DWT and WTT decompositions were applied to FTIR spectra of medicinal plant extracts as feature extraction techniques. Different configurations of preprocessing and decomposition parameters were tested for influence on the results of classification by Linear Discriminant Analysis and Logistic Regression and clustering by Hierarchical Clustering Analysis. While none of the used processing parameter combinations performed well on LDA, for LR and HAC DWT/WTT approaches showed high positive effects. In clustering, WTT decomposition demonstrated promising results as a part of processing pipeline. Overall, as with many other ML application, fine tuning of hyperparameters played important role in achieving better results.

Even though IR data was collected with no strict protocol (on either ratio of KBr or particle size before pressing mixtures into tablets), results showed that reasonable processing can make up for significant portion of such distortions. This is especially important in real-life routine applications where data is prone to be distorted due to many factors that are not easy to take into account.

Preprocessing and, co-dependently, feature extraction are vital to achieve high levels of classification accuracy and get meaningful insights from clustering. Various ways of handling raw data can either enhance or bury relevant chemical information, contained in data.

Robust and efficient approaches for quality control of herbal medicines continue to be vigorously sought after. Only established and continuously verified protocols for standardization and routine analysis of herbal preparations can allow many valid practices from CAM to be correctly assessed through clinical studies and be eventually integrated into modern Western pharmacology and medicine. Nevertheless, lack of explicit descriptions for data handling techniques used can make it hard to compare results obtained by different groups. The problem is further complicated by unavailability of raw data. And even though the amount of chemical data, cumulatively acquired through recent decades of plant research is massive and could potentially help advance the field, no major changes can be spotted as of now. In the long run, it can be rewarding to encourage and support practices of providing open access to raw scientific data when publishing new findings.

## Acknowledgement

This work is supported by Skoltech NGP Program (Skoltech-MIT joint project) and by the Ministry of Education and Science of the Russian Federation under grant 14.756.31.0001.

## Appendix A: Classification and clustering algorithms

*Linear discriminant analysis*

Linear discriminant analysis (LDA) can be defined as a direct approximation of Bayesian classifier with normaly distributed continuous variables. According to the Bayes theorem, conditional probability of observed sample $x \in \mathbb{R}^n$ to be a representative of class $s = \overline{1, S}$ is expressed as

$$P(y = s | X = x) = \frac{P(y=s)P(X=x|y=s)}{\sum_{j=1}^{S} P(y=j)P(X=x|y=j)} \qquad (6)$$

$$P(X = x | y = s) = \left((2\pi)^{n/2} \det(\Sigma_s)\right)^{-1} \exp\left(-\frac{1}{2}(x - \mu_s)^T \Sigma_s^{-1}(x - \mu_s)\right) \qquad (7)$$

where $\mu_s$ is a mean of $s$-th class, and all covariance matrices $\Sigma_s$ here are assumed to be equal, $\Sigma_s = \Sigma$. It may be shown that $s$ which maximizing expression $\left(\Sigma^{-1}x, \mu_s\right) - \frac{1}{2}\left(\Sigma^{-1}\mu_s, \mu_s\right) + \ln P(y = s)$ also maximizes the log-likelihood of the $P(y = s | X = x)$ what means that $s$ is the most probable prediction for a given sample $x$. Theoretical quantities $\mu_s$, $\Sigma$ are estimated from samples, and estimation of $P(y = s)$ is a simple occurrence frequency of class representatives in a training set. More complete explanation of LDA can be found in [45; 46].

*Logistic regression*

Basic logistic regression model is designed for binary classification problem and can be seen as a special case of generalized linear model. Multilabel case may be covered in different ways, for instance, one may train $S$ binary classifiers according to "one versus rest" strategy and select the label with maximal output among all $S$ classifiers. Binary logistic regression model uses the following hypothesis on the dependent variable:

$$y_k = h(x_k) = \theta\left(\sum_{i=1}^{n}\omega_i x_{ki}\right) = \theta(w^T x_k),$$ (8)

where $\theta(z) = \tanh(z)$ ( or $\theta(z) = (1 + e^{-z})^{-1}$, if $y_k \in \{0, 1\}$ ) is a sigmoid function, $w \in \mathbb{R}^n$ is a vector of parameters, $x_k \in \mathbb{R}^n$ is an input sample, $k = \overline{1, m}$.

In a logistic regression model one makes an attempt to estimate posterior probabilities using the above-mentioned hypothesis. Parameters $w$ are computed in order to minimize cross-entropy loss which is widely used to measure error between predicted output and true value of dependent variable:

$$w = \arg\min_{w}\left[\frac{1}{m}\sum_{k=1}^{m}\ln(1 + e^{-y_k \cdot (w, x_k)}) + \lambda\|w\|_p^2\right],$$ (9)

$(x_k, y_k)$ - $k$-th sample and its label, $x_k \in \mathbb{R}^n$, $y_k \in \{-1, +1\}$, $w \in \mathbb{R}^n$ - vector of parameters to estimate. The second additive component is a regularization term: it penalizes high values of $w$ with weight $\lambda$. Widely used types of regularizers are $l_1$ ($p = 1$, Lasso logistic regression) and $l_2$ ($p = 2$, Ridge logistic regression) ones. Detailed explanation of the algorithm can be found in [45–47].

*Hierarchical agglomerative clustering*

Clustering task in a general form may be formulated as follows: divide the samples by means of a specified distance function into such non-overlapping subsets that objects inside each subset (cluster) are as close as possible to each other and as far apart from members of other clusters. Instead of working with samples as in classical k-means approach, pairwise object-to-object similarities (or distances) may be utilized to perform clustering. One of clustering algorithms based on measuring certain pair-wise metrics for a set of objects is hierarchical agglomerative clustering. This clustering algorithm does not require any priors on the number of clusters. The only two things to be specified are metric of dissimilarities between objects and linkage that allows to recompute distances between merged objects or clusters. Initially, each object is associated with a single cluster. On further steps two currently closest clusters were merged into a new one, and the process is stopped with single remaining cluster which contains all objects from original set. Merging process may be visualized as a binary tree called dendrogram. Its structure may give insights into structure of the original dataset; however, this is a rather exploratory technique, and drawing conclusions based only on dendrogram plots should be avoided [45; 46].

Table 1: List of plant species used in the study.

| Species | Quantity |
| --- | --- |
| Linum usitatissimum | 12 |
| Glycyrrhiza glabra | 11 |
| Arctium lappa | 12 |
| Silybum marianum | 11 |
| Anethum graveolens | 14 |
| Inula helenium | 9 |
| Valeriana officinalis | 11 |

Table 2: Comparative results for best performing models on Linear Discriminant Analysis (mean of 25 times repeated 4-fold CV).

| Feature space | | Accuracy | | | F$_1$ (weighted) | | |
|---|---|---|---|---|---|---|---|
| | | $f$ | $f'$ | $f''$ | $f$ | $f'$ | $f''$ |
| original | (train) | 1.000 | 0.982 | 0.943 | 1.000 | 0.982 | 0.944 |
| | (test) | 0.833 | 0.849 | 0.800 | 0.827 | 0.841 | 0.786 |
| DWT (thr) | (train) | 0.991 | 0.948 | 0.955 | 0.991 | 0.949 | 0.956 |
| | (test) | 0.867 | 0.798 | 0.802 | 0.860 | 0.787 | 0.788 |
| DWT (sign) | (train) | 0.902 | 0.851 | 0.851 | 0.902 | 0.854 | 0.854 |
| | (test) | 0.861 | 0.871 | 0.871 | 0.856 | 0.865 | 0.865 |
| WTT (thr) | (train) | 0.982 | 0.963 | 0.943 | 0.982 | 0.963 | 0.944 |
| | (test) | 0.858 | 0.838 | 0.803 | 0.849 | 0.829 | 0.790 |
| WTT (sign) | (train) | 0.960 | 0.844 | 0.794 | 0.960 | 0.847 | 0.799 |
| | (test) | 0.812 | 0.853 | 0.770 | 0.803 | 0.848 | 0.760 |

Table 3: Comparative results for best performing models on Logistic Regression (mean of 25 times repeated 4-fold CV).

| Feature space | | Accuracy | | | F$_1$ (weighted) | | |
|---|---|---|---|---|---|---|---|
| | | $f$ | $f'$ | $f''$ | $f$ | $f'$ | $f''$ |
| original | (train) | 0.884 | 0.987 | 0.986 | 0.882 | 0.987 | 0.986 |
| | (test) | 0.723 | 0.946 | 0.885 | 0.711 | 0.944 | 0.877 |
| DWT | (train) | 1.000 | 1.000 | 1.000 | 1.000 | 1.000 | 1.000 |
| | (test) | 0.965 | 0.944 | 0.956 | 0.963 | 0.941 | 0.954 |
| WTT | (train) | 1.000 | 1.000 | 1.000 | 1.000 | 1.000 | 1.000 |
| | (test) | 0.969 | 0.959 | 0.897 | 0.968 | 0.957 | 0.894 |

Table 4: Comparative results for best performing models on Agglomerative Clustering (all samples were used; scored on precise number of classes).

| Score | Feature space | | |
|---|---|---|---|
| | original | DWT | WTT |
| | $f/f'/f''$ | $f/f'/f''$ | $f/f'/f''$ |
| Adjusted Rand index | 0.067/0.263/0.132 | 0.487/0.292/0.147 | 0.501/0.387/0.299 |
| Adjusted Mutual Information | 0.125/0.370/0.264 | 0.616/0.400/0.225 | 0.622/0.516/0.439 |
| Fowlkes-Mallows score | 0.229/0.400/0.369 | 0.573/0.470/0.283 | 0.582/0.474/0.414 |

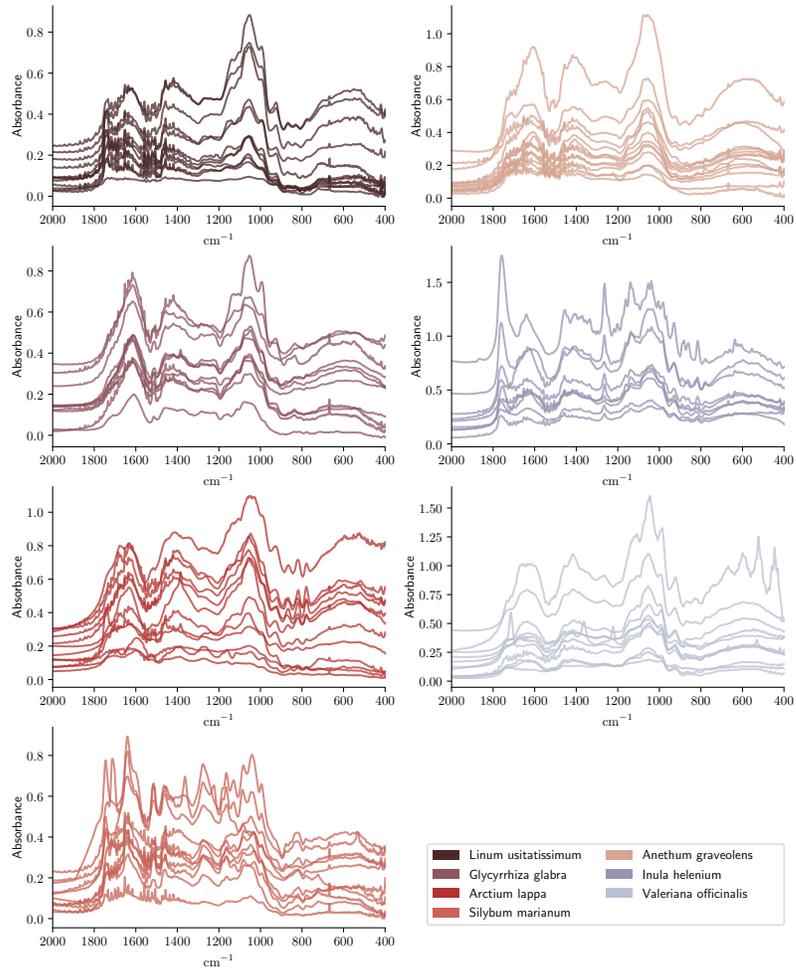

Figure 1: Fourier-transform infra-red spectra used in the study.

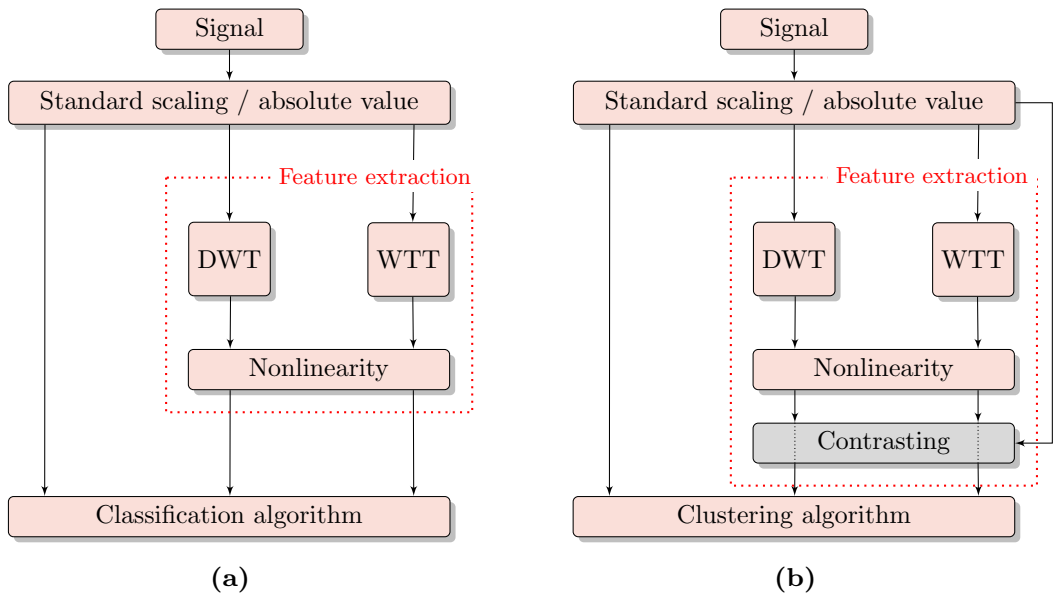

Figure 2: Schematic representation of general pipelines used in the study. (a): classification pipeline; (b) clustering pipeline; silver block ("contrasting") may be either used or ignored.

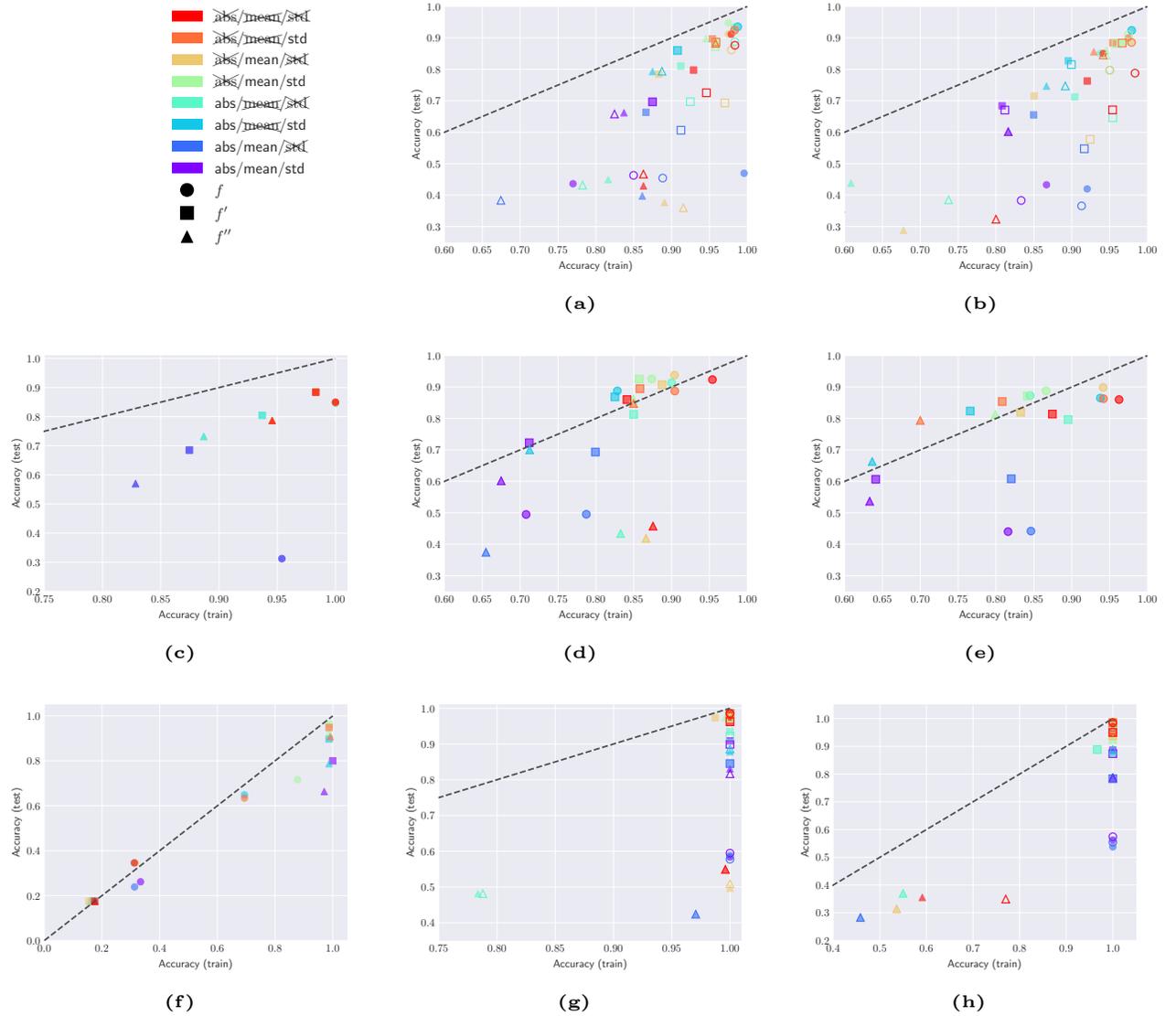

Figure 3: Visualization of accuracy values achieved on train and test parts of data for best performing models depending on different preprocessing steps (mean of 4-fold CV). (a) LDA, DWT with hard/soft thresholding; (b) LDA, WTT with hard/soft thresholding; (c) LDA, without decomposition; (d) LDA, DWT hard/soft thresholding followed by taking a signum; (e) LDA, DWT hard/soft thresholding followed by taking a signum; (f) LR, without decomposition; (g) LR, DWT hard/soft threshold followed by taking a signum; (h) LR, WTT hard/soft thresholding followed by taking a signum. In (a), (b), (d), (e), (g), (h) filled shapes correspond to hard thresholding, empty shapes - to soft thresholding.

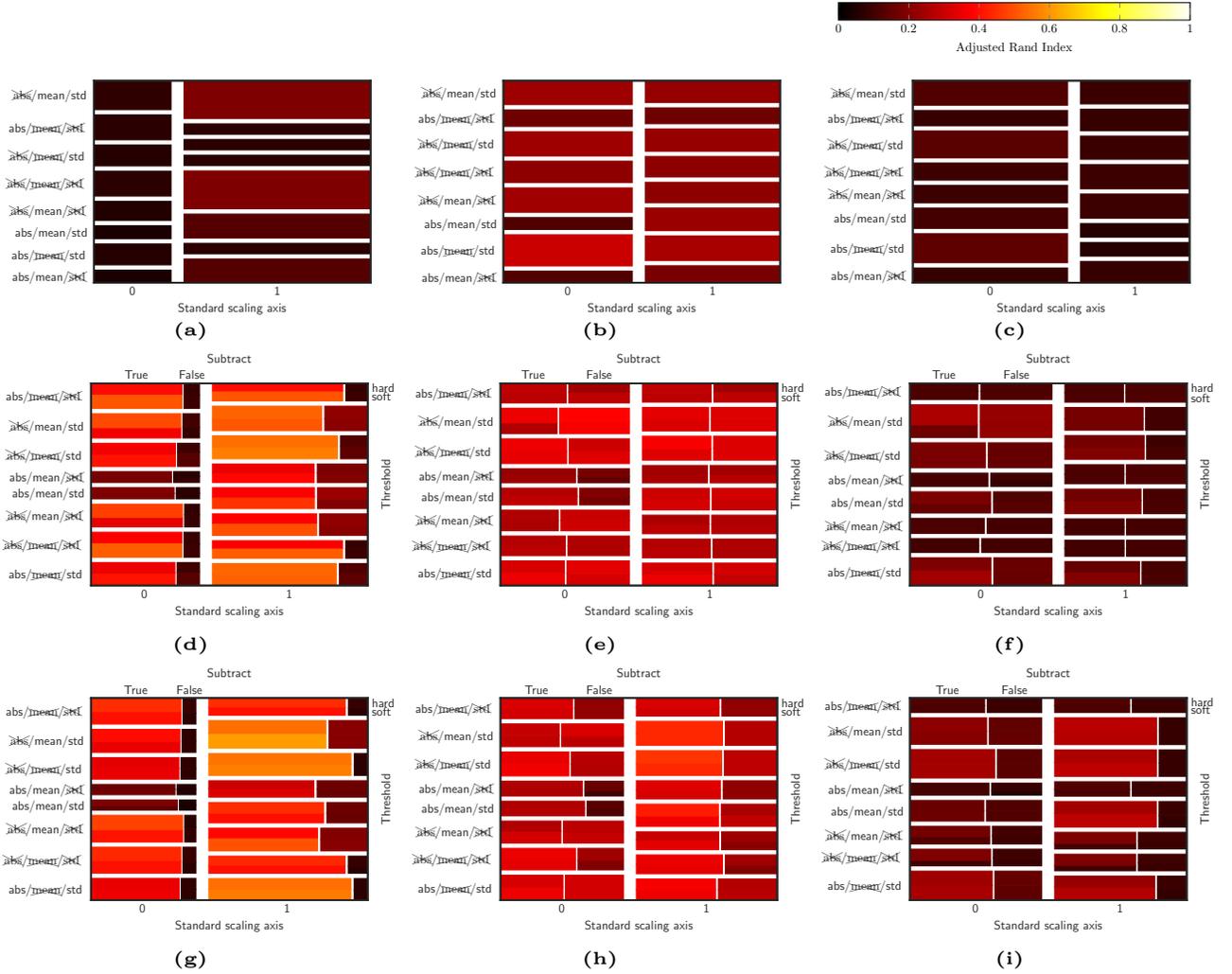

Figure 4: Influence of different parameter configurations on clustering results by means of adjusted Rand score. (a), (b), (c) - clustering without DWT/WTT decomposition, using original signal and its 1st and 2nd derivatives respectively; (d), (e), (f) - clustering with best performing DWT decompositions, using original signal and its 1st and 2nd derivatives respectively; (g), (h), (i) - clustering with best performing WTT decomposition using original signal and its 1st and 2nd derivatives respectively. Results were computed as mean of 4-fold CV-like scheme.

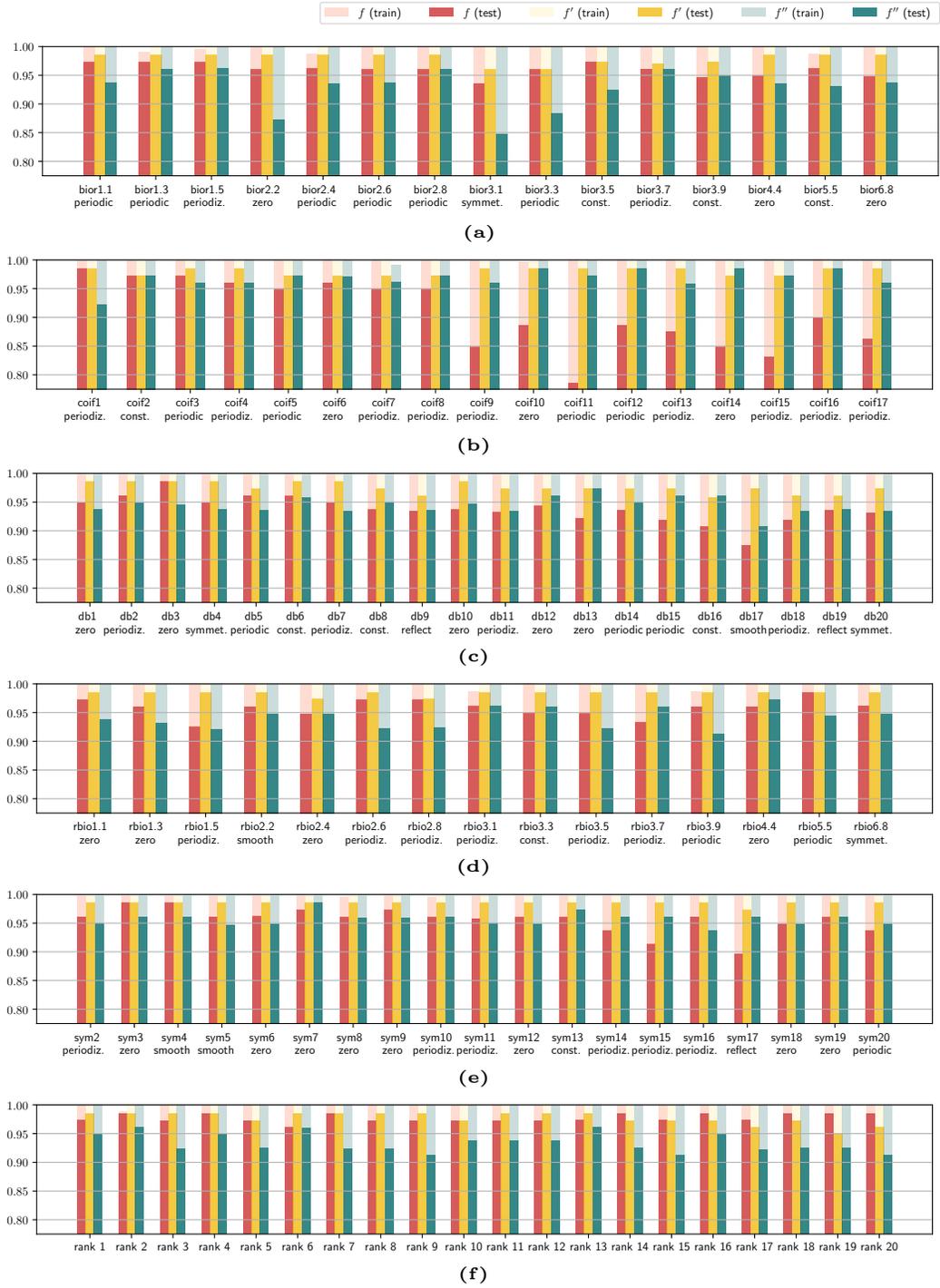

Figure 5: Performance of logistic regression depending on DWT/WTT decompositions and different values of hyperparameters. (a) biorthogonal wavelets; (b) coiflets; (c) Daubechies wavelets; (d) reverse biorthogonal wavelets; (e) symlets; (f) WTT with different ranks. In (a)–(e) padding was chosen to maximize accuracy for given wavelet on both original and derivative spectra with other parameters being individual for each type of spectra. Results were obtained by taking mean of accuracy values in 4-fold CV.

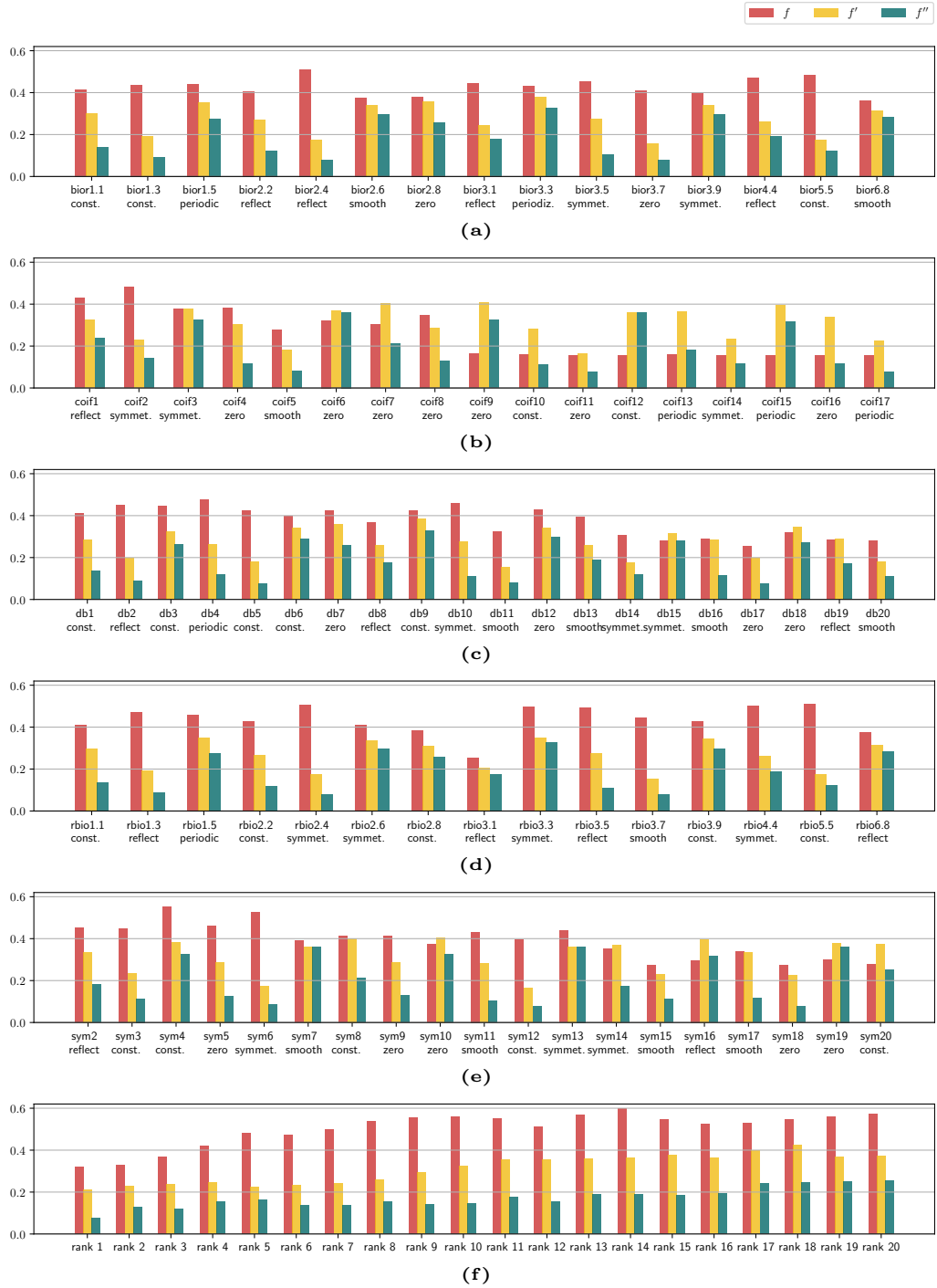

Figure 6: Performance of agglomerative clustering in terms of adjusted Rand index (ARI) for 5 DWT families and WTT decomposition. (a) biorthogonal wavelets; (b) coiflets; (c) Daubechies wavelets; (d) reverse biorthogonal wavelets; (e) symlets; (f) WTT with different ranks. In (a)–(e) padding was chosen to maximize ARI for given wavelet on both original and derivative spectra with other parameters being individual for each type of spectra.

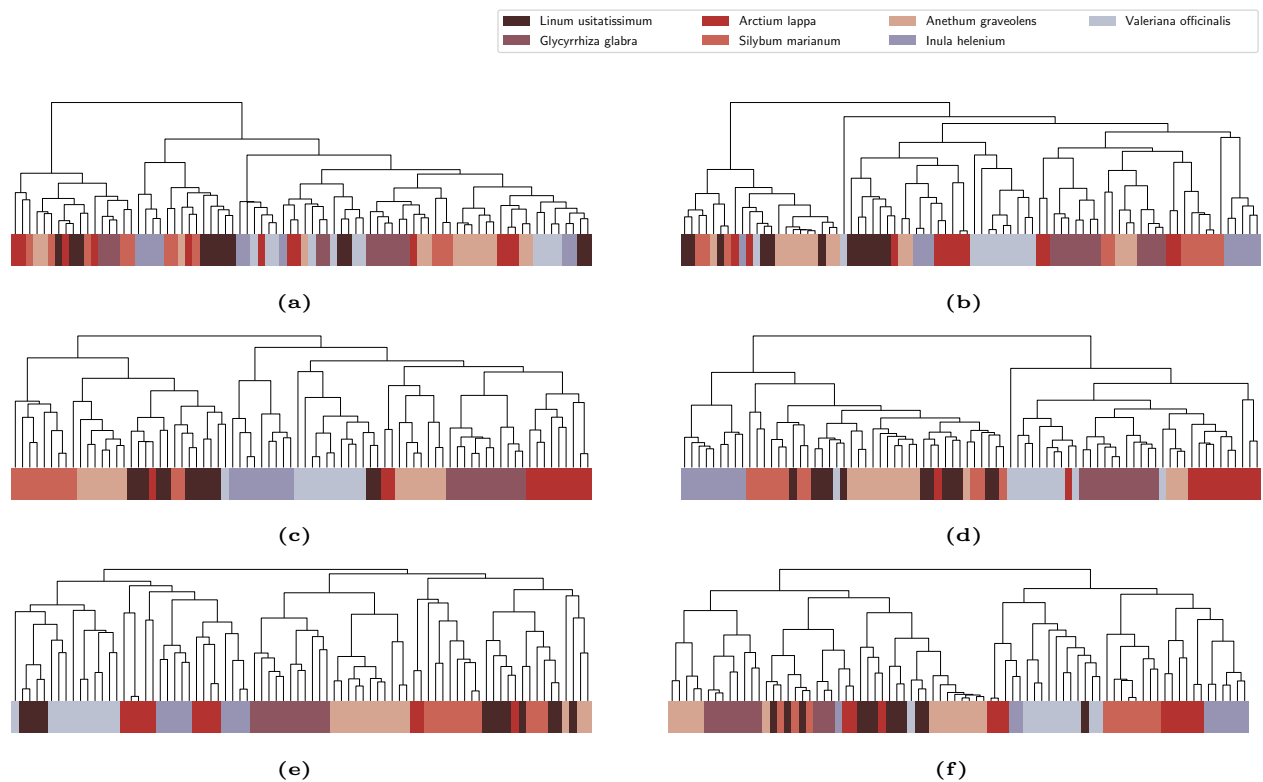

Figure 7: Dendrogram plots computed as the result of hierarchical agglomerative clustering of all available samples. (a) clustering without DWT/WTT processing, oriiginal signal; (b) clustering without DWT/WTT processing, first derivative; (c) clustering with best performing DWT, original signal; (d) clustering with best performing DWT, first derivative; (e) clustering with best performing WTT, original signal; (f) clustering with best performing WTT, first derivative.